\documentclass[showpacs,amsmath,twocolumn,showkeys,pra]{revtex4}
\usepackage{dcolumn}    
\usepackage{ifpdf}
\usepackage{amssymb,lineno,amsfonts}
\usepackage{graphicx}   
\usepackage{dcolumn}    
\usepackage{bm}         
\usepackage{bbm}
\usepackage{mathrsfs}
\usepackage{upgreek}
\usepackage{mathtools}
\usepackage{epstopdf}
\usepackage{setspace}
\usepackage{hyperref}
\usepackage{float}
\usepackage{natbib}
\usepackage[usenames,dvipsnames]{xcolor}

\newcommand{\ket}[1]{\vert{#1}\rangle}
\newcommand{\bra}[1]{\langle{#1}\vert}
\newcommand{\outpr}[2]{\vert{#1}\rangle\langle{#2}\vert}

\definecolor{med-blue}{RGB}{25,25,112}
\hypersetup{colorlinks, linkcolor={red},citecolor={blue}, urlcolor={MidnightBlue}}

\begin{document}

\title{Inversion of moments to retrieve joint probabilities in quantum sequential measurements}
\author{H. S. Karthik} 
\affiliation{Raman Research Institute, Bangalore 560 080, India} 
\author{Hemant Katiyar} 
\affiliation{Department of Physics and NMR Research Center,
Indian Institute of Science Education and Research, Pune 411008, India}
\author{Abhishek Shukla}
\affiliation{Department of Physics and NMR Research Center,
Indian Institute of Science Education and Research, Pune 411008, India}
\author{A. R. Usha Devi} 
\affiliation{Department of Physics, Bangalore University, 
Bangalore-560 056, India}
\affiliation{Inspire Institute Inc., Alexandria, Virginia, 22303, USA.}
\email{arutth@rediffmail.com}
\author{A. K. Rajagopal} 
 \affiliation{Inspire Institute Inc., Alexandria, Virginia, 22303, USA.}
\affiliation{Harish-Chandra Research Institute, Chhatnag Road, Jhunsi, Allahabad 211 019, India.}
\author{T. S. Mahesh} 
\affiliation{Department of Physics and NMR Research Center,
Indian Institute of Science Education and Research, Pune 411008, India}

\begin{abstract}
A sequence of  moments obtained from statistical trials encodes a classical probability distribution. However, it is well-known that an incompatible set of moments arise in the quantum scenario, when correlation outcomes associated with measurements on spatially separated entangled states are considered. This feature viz., the incompatibility of moments with a joint probability distribution is reflected in the violation of Bell inequalities. Here, we focus on sequential measurements on a single quantum system and investigate if moments and joint probabilities are compatible with each other. By considering sequential measurement of a dichotomic dynamical observable at three different time intervals, we explicitly demonstrate that the moments and the probabilities are inconsistent with each other.  Experimental results using a nuclear magnetic resonance (NMR) system are reported here to corroborate these theoretical observations viz., the incompatibility of the three-time joint probabilties with those extracted from the moment sequence when sequential measurements on a single qubit system are considered.
\end{abstract}

\maketitle

\section{Introduction}

The issue of determining a probability distribution uniquely in terms of its moment sequence -- known as {\em classical moment problem} -- has been developed for more than 100 years~\cite{Tamarkin, Akhiezer}. In the case of discrete distributions with the associated random variables taking finite values, moments faithfully capture the essence of the probabilities i.e., the probability distribution is moment determinate~\cite{Shi}. 

In the special case of classical random variables $X_i$ assuming dichotomic values $x_i=\pm 1$, it is easy to see that the sequence of  moments~\cite{note1} 
$\mu_{n_1\,n_2\,\ldots n_k}=\langle X_1^{n_1}X_2^{n_2}\ldots X_k^{n_k}\rangle=\displaystyle{\sum_{x_1,x_2,\ldots\, x_k=\pm 1}}\, x^{n_1}_1\, x^{n_2}_2\ldots\,x^{n_k}_k\, P(x_1,x_2,\ldots , x_k)\ ,$             
where $n_1,n_2,\ldots, n_k=0,1,$  can be readily inverted to obtain the joint probabilities  $P(x_1,x_2,\ldots , x_k)$ uniquely.  More explicitly,  the joint probabilities $P(x_1,x_2,\ldots , x_k)$ are given in terms of the $2^k$ moments  $\mu_{n_1\, n_2\, \ldots n_k}, n_1,n_2,\ldots n_k=0,1$ as, 
\begin{eqnarray}
\label{1}
 P(x_1,x_2,\ldots, x_k)=\frac{1}{2^k}\sum_{n_1,\ldots n_k=0, 1}\, x_1^{n_1}x_2^{n_2}\ldots x_k^{n_k}\, \mu_{n_1\, \ldots n_k}\nonumber \\ 
 =\frac{1}{2^k}\, \sum_{n_1,\ldots, n_k=0, 1}\, x_1^{n_1}x_2^{n_2}\ldots x_k^{n_k}\, \langle X_1^{n_1}X_2^{n_2}\ldots X_2^{n_k}\rangle. \nonumber \\
\end{eqnarray}

Does this feature prevail in the quantum scenario? This results in a negative answer as it is wellknown that the moments associated with measurement outcomes on spatially separated parties are not compatible with the joint probability distribution. This feature reflects itself in the violation of Bell inequalities. In this paper we investigate whether moment-indeterminacy persists when we focus on sequential measurements on a single quantum system. We show that the discrete joint probabilities originating in the sequential measurement of a  single qubit dichotomic observable $\hat{X}(t_i)=\hat{X}_i$ at  different time intervals are not consistent  with the ones reconstructed from the moments.  More explicitly, considering  sequential measurements of $\hat{X}_1$, $\hat{X}_2$, $\hat{X}_3$, we reconstruct the trivariate joint probabilties $P_{\mu}(x_1,x_2,x_3)$  based on the set of eight moments $\{\langle \hat{X}_1\rangle, \langle \hat{X}_2\rangle,  \langle \hat{X}_3\rangle, \langle \hat{X}_1\, \hat{X}_2\rangle, \langle \hat{X}_2\, \hat{X}_3\rangle, \langle \hat{X}_1\, \hat{X}_3\rangle, \langle \hat{X}_1\, \hat{X}_2\, \hat{X}_3\rangle \}$ and prove that they do not agree with the three-time  joint probabilities (TTJP) $P_d(x_1,x_2,x_3)$ evaluated directly based on the correlation outcomes in the sequential measurement of all the three observables. Interestingly, the moments and TTJP can be independently extracted experimentally in NMR system -- demonstrating the difference between moment inverted three time probabilities with the ones directly drawn from experiment, in agreement with theory. For obtaining TTJP directly we use the procedure of Ref.~\cite{nmr_elgi} and for extracting moments we extend the Moussa protocol~\cite{Moussa} to a set of non-commutating observables. The specifics are given in the experimental section.

Disagreement between moment inverted joint probabilities with the ones based on measurement outcomes in turn reflects the inherent inconsistency  that the family of all marginal probabilities do not arise from the grand joint probabilities. The non-existence of a legitimate grand joint probability distribution, consistent with the set of all pairwise marginals is attributed to be the common origin of a wide range of no-go theorems on non-contextuality, locality and macrorealism in the foundations of quantum theory~\cite{Fine, Bell, KS, Peres, Mermin, LG, UKSR, TP2013}. The absence of a valid grand joint probability distribution in the sequential measurement on a single quantum system is brought out here in terms of its mismatch with moment sequence. 

We organize the paper as follows. In Sec.~II we begin  with a discussion on moment inversion to obtain joint probabilities of three classical random variables assuming dichotomic values $\pm 1$. We proceed in Sec.~III to study the quantum scenario with the help of a specific example of sequential measurements of dichotomic observable at three different times on a spin-1/2 system. We show that the TTJP constructed from  eight moments do not agree with those originated from the measurement outcomes. Sec.~IV is devoted to report experimental results with NMR implementation  on an ensemble of  spin-1/2 nuclei, demonstrating that moment constructed TTJP do not agree with those directly extracted. Section V has concluding remarks.            
    
\section{Reconstruction of joint probability of classical dichotomic random variables from moments}     

Let $X$ denote a dichotomic random variable with outcomes $x=\pm 1$. The moments associated with statistical outcomes involving the variable $X$  are given by  
$\mu_n=\langle X^n\rangle=\sum_{x=\pm 1} x^n\, P(x),\ n=0,1,2,3, \ldots$, where $0\leq P(x=\pm 1)\leq 1; \sum_{x=\pm 1}P(x)=1$ are the corresponding probabilities.  Given the moments $\mu_0$ and $\mu_1$ from a statistical trial, 
one can readily  obtain the probability mass function: 
\begin{eqnarray*}
 P(1)&=&\frac{1}{2}(\mu_0+\mu_1)=\frac{1}{2}(1+\mu_1)\\ 
 P(-1)&=& \frac{1}{2}(\mu_0-\mu_1)=\frac{1}{2}(1-\mu_1), 
\end{eqnarray*}  
i.e., moments  determine the probabilities uniquely. 

In the case of two dichotomic random variables $X_1$, $X_2$, the moments  
$\mu_{n_1,n_2}=\langle X_1^{n_1}\, X_2^{n_2}\rangle=\displaystyle\sum_{x_1=\pm 1, x_2=\pm 1}\, x_1^{n_1}\, x_2^{n_2}\, P(x_1,x_2),\ n_1,n_2=0,1\ldots$ encode the bivariate probabilities $P(x_1,x_2)$. Explicitly, 
\begin{widetext}
\begin{eqnarray} 
\mu_{00}&=& \sum_{x_1, x_2=\pm 1}\, P(x_1,x_2)=P(1,1)+P(1,-1)+P(-1,1)+P(-1,-1)=1, \nonumber \\ 
\mu_{10}&=& \sum_{x_1, x_2=\pm 1}\, x_1\, P(x_1,x_2)= \sum_{x_1=\pm 1}\, x_1\, \,P(x_1), \nonumber \\
&=& P(1,1)+P(1,-1)-P(-1,1)-P(-1,-1) \nonumber \\ 
\mu_{01}&=& \sum_{x_1, x_2=\pm 1} x_2\, P(x_1,x_2)= \sum_{x_2}\, x_2 \,P(x_2) \nonumber \\
&=& P(1,1)-P(1,-1)+P(-1,1)-P(-1,-1) \nonumber \\ 
\mu_{11}&=& \sum_{x_1, x_2=\pm 1}\, x_1\, x_2\, P(x_1,x_2)=P(1,1)-P(1,-1)-P(-1,1)+P(-1,-1). 
\end{eqnarray}  
\end{widetext}
Note that the  moments $\mu_{10}$, $\mu_{01}$  involve the marginal probabilities $P(x_1)=\sum_{x_2=\pm 1}\, P(x_1,x_2)$, $P(x_2)=\sum_{x_1=\pm 1}\, P(x_1,x_2)$ respectively and they could be evaluated based on statistical trials  drawn independently from the two random variables $X_1$ and $X_2$.  

Given the moments $\mu_{00}, \mu_{10}, \mu_{01}, \mu_{11}$ the reconstruction of  the probabilities $P(x_1,x_2)$ is straightforward:   
 \begin{eqnarray}
 P(x_1,x_2)&=& \frac{1}{4}\, \sum_{n_1, n_2=0,1}\, x_1^{n_1}x_2^{n_2}\,  \mu_{n_1\, n_2} \nonumber \\ 
 &=& \frac{1}{4}\, \sum_{n_1, n_2=0,1}\, x_1^{n_1}x_2^{n_2}\, \langle X_1^{n_1}\, X_2^{n_2}\rangle. 
\end{eqnarray}

Further, a reconstruction of trivariate joint probabilities $P(x_1,x_2,x_3)$ requires the following set of  eight moments: 
$\{\mu_{000}=1, \mu_{100}=\langle X_1\rangle,\ \mu_{010}=\langle X_2\rangle, \mu_{010}=\langle X_3\rangle, \mu_{110}=\langle X_1\, X_2\rangle, \mu_{011}=\langle X_2\, X_3\rangle, \mu_{101}=\langle X_1\, X_3\rangle,\ \mu_{111}=\langle X_1\, X_2\, X_3\rangle\}$. The probabilities are retrieved faithfully in terms of the eight moments as,    
\begin{eqnarray}
\label{MI}
P(x_1,x_2,x_3)= \frac{1}{8}\, \sum_{n_1, n_2, n_3=0,1}\, x_1^{n_1}x_2^{n_2}x_3^{n_3}\,  \mu_{n_1\, n_2\, n_3} \nonumber \\ 
=\frac{1}{8}\, \sum_{n_1, n_2, n_3=0,1}\, x_1^{n_1}x_2^{n_2}x_3^{n_3}\, \langle X_1^{n_1}\, X_2^{n_2}\, X^{n_3}_3\rangle. \nonumber \\
\end{eqnarray} 
It is implicit that the moments  $\mu_{100}, \mu_{010}, \mu_{001}$ are determined through independent statistical trials involving the random variables $X_1, X_2, X_3$ separately;  $\mu_{110}, \mu_{011}, \mu_{101}$ are obtained based on the correlation outcomes of $(X_1, X_2)$, $(X_2, X_3)$ and $(X_1, X_3)$ respectively. More specifically, in the classical probability setting there is a  tacit underlying assumption that the set of all marginal probabilities $P(x_1), P(x_2), P(x_3), P(x_1,x_2), P(x_2,x_3), P(x_1,x_3)$ are consistent with the trivariate joint probabilities $P(x_1,x_2,x_3)$. This underpinning does not get imprinted automatically in the quantum scenario. Suppose the observables $\hat{X}_1, \hat{X}_2, \hat{X}_3$ are non-commuting and we consider their sequential measurement.  The moments $\mu_{100}=\langle \hat{X}_1\rangle, \mu_{010}=\langle \hat{X}_2\rangle, \ \mu_{001}=\langle \hat{X}_3\rangle$ may be evaluated from the measurement outcomes of dichotomic observables $\hat{X}_1, \hat{X}_2, \hat{X}_3$ independently;  the correlated statistical outcomes in the sequential measurements of $(\hat{X}_1,\ \hat{X}_2)$, $(\hat{X}_2,\ \hat{X}_3)$ and $(\hat{X}_1,\ \hat{X}_3)$ allow one to extract the set of moments  $\mu_{110}=\langle \hat{X}_1\hat{X}_2\rangle, \ \mu_{011}=\langle \hat{X}_2\hat{X}_3\rangle, \mu_{101}=\langle\,\hat{X}_1\hat{X}_3\rangle$; further the moment $\mu_{111}=\langle \hat{X}_1\,\hat{X}_2\, \hat{X}_3\rangle $ is evaluated based on the correlation outcomes when all the three observables are measured sequentially. The joint probabilities $P_\mu(x_1,x_2,x_3)$ retrieved from the moments as given in (\ref{MI}) differ from   the ones evaluated directly in terms of the correlation outcomes in the sequential measurement of all the three observables . We illustrate this inconsistency appearing in the quantum setting in the next section.

\section{Quantum three-time joint probabilities and moment inversion}
 
Let us consider a spin-1/2 system, dynamical evolution of which is governed by the Hamiltonian 
\begin{equation}
\hat{H}=\frac{1}{2}\, \hbar\, \omega \sigma_x. 
\end{equation} 
We choose z-component of spin as our dynamical observable: 
\begin{eqnarray}
\hat{X}_i&=&\hat{X}(t_i)=\sigma_z(t_i) \nonumber \\
&=& \hat{U}^\dag(t_i)\, \sigma_z\, \hat{U}(t_i) \nonumber \\
&=&\sigma_z\, \cos\omega\, t_i+\sigma_y\, \sin\omega\, t_i,
\end{eqnarray} 
where  $\hat{U}(t_i)=e^{-i\,\sigma_x\,\omega\, t_i/2}=\hat{U}_i$,  
and consider sequential measurements of the observable $\hat{X}_i$ at three different times $t_1=0, t_2=\Delta t, t_3=2\, \Delta t$: 
\begin{eqnarray}
\hat{X}_1&=&\sigma_z\nonumber \\ 
 \hat{X}_2&=&\sigma_z(\Delta t)=\sigma_z\, \cos(\omega\Delta t)+\sigma_y\, \sin(\omega\Delta t) \nonumber \\ 
\hat{X}_3&=& \sigma_z(2\Delta t)=\sigma_z\, \cos(2\omega\Delta t)+\sigma_y\, \sin(2\omega\Delta t).
\end{eqnarray}
Note that these three operators are not commuting in general. 

The moments $\langle \hat{X}_1\rangle, \langle \hat{X}_2\rangle, \langle \hat{X}_3\rangle$ are readily evaluated to be  
\begin{eqnarray*}
\label{1mom}
 \mu_{100}&=&\langle \hat{X}_1\rangle = {\rm Tr}[\hat{\rho}_{\rm in}\, \sigma_z]=0,  \\ 
\mu_{010}&=&\langle \hat{X}_2\rangle={\rm Tr}[\hat{\rho}_{\rm in}\, \sigma_z(\Delta t)]=0,  \\ 
\mu_{001}&=& \langle \hat{X}_3\rangle={\rm Tr}[\hat{\rho}_{\rm in}\, \sigma_z (2\Delta t)]=0.
\end{eqnarray*}
when the system density matrix is prepared initially in a maximally mixed state $\hat{\rho}_{\rm in}=\mathbbm{1}/2$. The probabilities of outcomes $x_i=\pm 1$ in the completely random initial state are given by $P(x_i=\pm 1)={\rm Tr}[\hat{\rho}_{\rm in}\, \hat{\Pi}_{x_i}]=\frac{1}{2}$, where  $\hat{\Pi}_{x_i}=\vert x_i\rangle\langle x_i\vert$ is the projection operator corresponding to measurement of the observable $\hat{X}_i$. 

The two-time joint probabilities arising in the sequential measurements of the observables $\hat{X}_i, \hat{X}_{j},\ j>i$ are evaluated as follows. 
The measurement of the observable $\hat{X}_i$ yielding the outcome $x_i=\pm 1$ projects the the density operator to $\hat{\rho}_{x_i}=\frac{\hat{\Pi}_{x_i}\, \hat{\rho}_{\rm in}\, \hat{\Pi}_{x_i}}{{\rm Tr}[\hat{\rho}_{\rm in}\, \hat{\Pi}_{x_i}]}$. Further, a sequential measurement of  $\hat{X}_j$ leads to the  two-time joint probabilities  as, 

\begin{eqnarray}
\label{pxixj}
P(x_i,x_j)&=& P(x_i)\, P(x_j\vert x_i)\nonumber \\ 
&=& {\rm Tr}[\hat{\rho}_{\rm in}\, \hat{\Pi}_{x_i}]\ {\rm Tr}[\hat{\rho}_{x_i}\, \hat{\Pi}_{x_j}] \nonumber \\
&=&{\rm Tr}[\hat{\Pi}_{x_i}\,\hat{\rho}_{\rm in}\, \hat{\Pi}_{x_i} \hat{\Pi}_{x_j}] \nonumber \\
&=& \langle x_i\vert\, \hat{\rho}_{\rm in}\, \vert x_i\rangle\, \vert\langle x_i\vert x_j\rangle\vert^2   
\end{eqnarray}  
 
We evaluate the two-time joint probabilities associated with the sequential measurements of 
$(\hat{X}_1,\hat{X}_2)$, $(\hat{X}_2,\hat{X}_3)$, and $(\hat{X}_1,\hat{X}_3)$ explicitly: 
\begin{eqnarray}
P(x_1, x_2)=\frac{1}{4}\, [1+ x_1\, x_2\, \cos(\omega\Delta t)] \\
P(x_2, x_3)=\frac{1}{4}\, [1+ x_2\, x_3\, \cos(\omega\Delta t)] \\
P(x_1, x_3)=\frac{1}{4}\, [1+ x_1\, x_3\, \cos(2\omega\Delta t)].   
\end{eqnarray}
We then obtain two-time correlation moments as,
\begin{eqnarray}
\label{2mom}
\mu_{110}=\langle\, \hat{X}_1\hat{X}_2\rangle&=&\sum_{x_1,x_2=\pm 1}\, x_1\,x_2\, P(x_1, x_2) \nonumber \\
&=& \cos(\omega\Delta t) \\
\mu_{011}=\langle\, \hat{X}_2\hat{X}_3\rangle&=&\sum_{x_2,x_3=\pm 1}\, x_2\,x_3\, P(x_2, x_3) \nonumber \\
&=&\cos(\omega\Delta t) \\
\mu_{101}=\langle\, \hat{X}_1\hat{X}_3\rangle&=&\sum_{x_1,x_3=\pm 1}\, x_1\,x_3\, P(x_1, x_3) \nonumber \\
&=& \cos(2\, \omega\Delta t).
\end{eqnarray}
Further, the three-time joint probabilities $P(x_1,x_2,x_3)$ arising in the sequential measurements of $\hat{X}_1, \hat{X}_2$, followed by  $\hat{X}_3$ are given by
\begin{eqnarray}
P(x_1,x_2,x_3)&=& P(x_1)\, P(x_2\vert x_1)\, P(x_3\vert x_1,x_2)\nonumber \\ 
&=& {\rm Tr}[\hat{\rho}_{\rm in}\, \hat{\Pi}_{x_1}]\, {\rm Tr}[\hat{\rho}_{x_1}\, \hat{\Pi}_{x_2}]\, {\rm Tr}[\hat{\rho}_{x_2}\, \hat{\Pi}_{x_3}] \nonumber \\
\end{eqnarray} 
where $\hat{\rho}_{x_2}=\frac{\hat{\Pi}_{x_2}\, \hat{\rho}_{x_1}\, \hat{\Pi}_{x_2}}{{\rm Tr}[\hat{\rho}_{x_1}\, \hat{\Pi}_{x_2}]}.$ 
We obtain, 
\begin{eqnarray}
\label{dirTTJP}
P(x_1,x_2,x_3)&=&{\rm Tr}[\hat{\Pi}_{x_2}\, \hat{\Pi}_{x_1}\, \hat{\rho}_{\rm in}\, \hat{\Pi}_{x_1}\, \hat{\Pi}_{x_2}\, \hat{\Pi}_{x_3}]\nonumber \\ 
 &=&  \langle x_1\vert\, \hat{\rho}_{\rm in}\, \vert x_1\rangle\, \vert\langle x_1\vert x_2\rangle\vert^2\, \vert\langle x_2\vert x_3\rangle\vert^2 \nonumber \\
 &=& \frac{P(x_1, x_2)\, P(x_2, x_3)}{\langle x_2\vert\, \hat{\rho}_{\rm in}\, \vert x_2\rangle}\nonumber \\ 
 &=& \frac{P(x_1, x_2)\, P(x_2, x_3)}{P(x_2)} 
 \end{eqnarray}
where in the third line of (\ref{dirTTJP}) we have used (\ref{pxixj}).   

The three-time correlation moment is evaluated to be, 
\begin{eqnarray}
\label{3mom}
\mu_{111}=\langle \hat{X}_1\, \hat{X}_2\, \hat{X}_3\rangle&=& \sum_{x_1,x_2,x_3=\pm 1}\, x_1\, x_2\,x_3\, P(x_1, x_2, x_3)\nonumber \\
&=& 0.
\end{eqnarray}

From the set of eight moments (\ref{1mom}), (\ref{2mom}) and (\ref{3mom}), we construct the TTJP (see (\ref{MI})) as,  
\begin{widetext}
\begin{eqnarray}
\label{mip}
P_{\mu}(1,1,1)&=&\frac{1}{8}\ [1+2\cos(\omega\Delta t)+\cos(2\omega\Delta t)] = P_{\mu}(-1,-1,-1), \nonumber \\ 
P_{\mu}(-1,1,1)&=&\frac{1}{8}\ [1-\cos(2\omega\Delta t)]=P_{\mu}(-1,-1,1) = P_{\mu}(1,1,-1) = P_{\mu}(1,-1,-1),  \\
P_{\mu}(1,-1,1)&=&\frac{1}{8}\ [1-2\cos(\omega\Delta t)+\cos(2\omega\Delta t)] = P_{\mu}(-1,1,-1). \nonumber
\end{eqnarray}
On the other hand, the three dichotomic variable quantum probabilities $P(x_1,x_2, x_3)$ evaluated  directly are given by,  
\begin{eqnarray}
\label{pdirect}
P_{d}(1,1,1)&=& \frac{1}{8}\, [1+\cos(\omega\, \Delta t)]^2=P_{d}(-1,-1,-1), \nonumber \\
P_{d}(-1,1,1)&=& \frac{1}{8}\, [1-\cos^2(\omega\, \Delta t)] = P_{d}(-1,-1,1)=P_{d}(1,1,-1)=P_{d}(1,-1,-1), \\  
P_{d}(1,-1,1)&=& \frac{1}{8}\, [1-\cos(\omega\, \Delta t)]^2=P_{d}(-1,1,-1).  \nonumber 
\end{eqnarray} 
\end{widetext}

Clearly, there is no agreement between the moment inverted TTJP (\ref{mip}) and the ones of (\ref{pdirect}) directly evaluated. In other words, the TTJP  realized in a sequential measurement are not invertible in terms of the moments -- which in turn reflects the incompatibility of the set of all marginal probabilities with the grand joint probabilities  $P_d(x_1,x_2,x_3).$ In fact, it may be explicitly verified that $P(x_1,x_3)\neq \sum_{x_2=\pm 1}\, P_d(x_1,x_2,x_3).$ 
Moment-indeterminacy points towards the absence of a valid grand probability distribution consistent with all the marginals.   

The TTJP and moments can be independently extracted experimentally using NMR methods on an ensemble of spin-1/2 nuclei. The experimental approach and results are reported in the next section.   
  
\section{Experiment}

The projection operators at time $t=0$ ($ \hat{X}_1 = \sigma_z $) are 
$\{\hat{\Pi}_{x^0_i} =  \outpr{x^0_i}{x^0_i} \}_{x^0_i= 0,1}$.
This measurement basis is rotating under the unitary $\hat{U}_i$, resulting in time dependent basis given by,
$ \hat{\Pi}_{x_i^t}= \hat{U}^\dag_i\, \hat{\Pi}_{x^0_i} \hat{U}_i$.
While doing experiments it is convenient to perform the measurement in the computational basis as 
compared to the time dependent basis. This can be done as follows: We can expand the measurement on an 
instantaneous state $\rho(t_i)$ as,
$\hat{\Pi}_{x_i^t} \hat{\rho}(t_i) \hat{\Pi}_{x_i^t} = \hat{U}_i^\dagger \hat{\Pi}_{x_i^0} \left( \hat{U}_i \hat{\rho}(t_i) \hat{U}_i^\dagger \right) \Pi_{x_i^0} U_i$. Thus, measuring in time dependent basis is equivalent to  evolving the state under the unitary $\hat{U}_i$, followed by measuring in the computational basis and lastly evolving under the unitary $\hat{U}_i^\dagger$. 
\begin{figure}
\centering
\includegraphics[width=8cm]{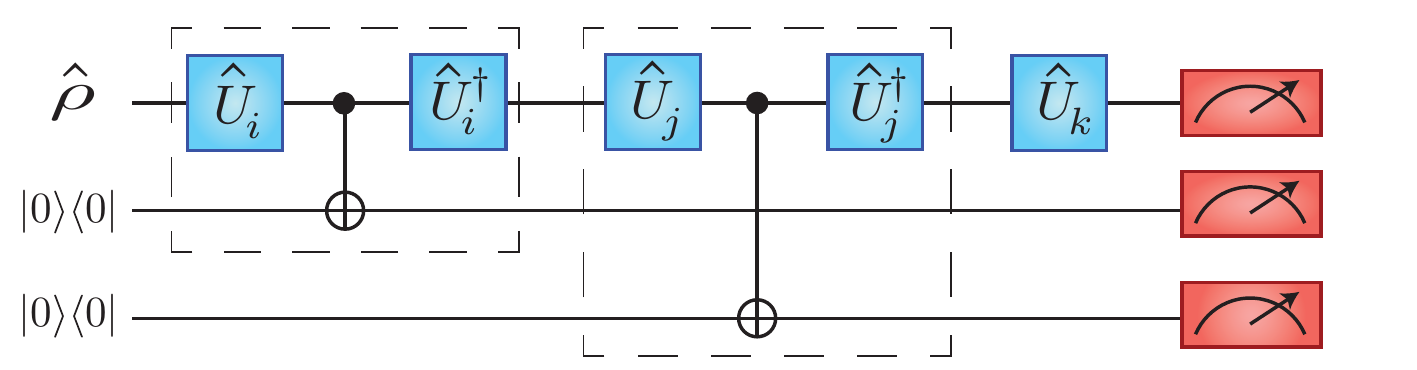}
\caption{Circuit for finding three-time probability. Grouped gates represent
measurement in rotated basis and controlled gates can be CNOT or anti-CNOT
as explained in the text.}
\label{ttjp}
\end{figure}

The probabilities of measurement outcomes can be encoded onto the ancilla qubits with the help of CNOT (or anti-CNOT) gate.
To see this property consider a one qubit general state (for system) and an ancilla in the state
$ \outpr{0}{0} $, then the CNOT gate encodes the probabilities as follows
\begin{eqnarray}
& \left( p_0 \outpr{0}{0} + p_1 \outpr{1}{1} + a \outpr{1}{0} + a^\dagger \outpr{0}{1} \right) _S  
\otimes \outpr{0}{0}_A & \nonumber \\
&\downarrow \mathrm{CNOT}& \nonumber \\
& \outpr{0}{0}_S \otimes p_0 \outpr{0}{0}_A 
+ \outpr{1}{1}_S \otimes p_1 \outpr{1}{1}_A & \nonumber \\
& + \outpr{1}{0}_S \otimes a \outpr{1}{0}_A
+ \outpr{0}{1}_S \otimes a^\dagger \outpr{0}{1}_A.& \nonumber
\end{eqnarray}
Now measuring the diagonal terms of the ancilla qubit, we can retrieve $ p_0 $ and $ p_1 $.
\begin{figure}[b]
\centering
\includegraphics[width=8.4cm]{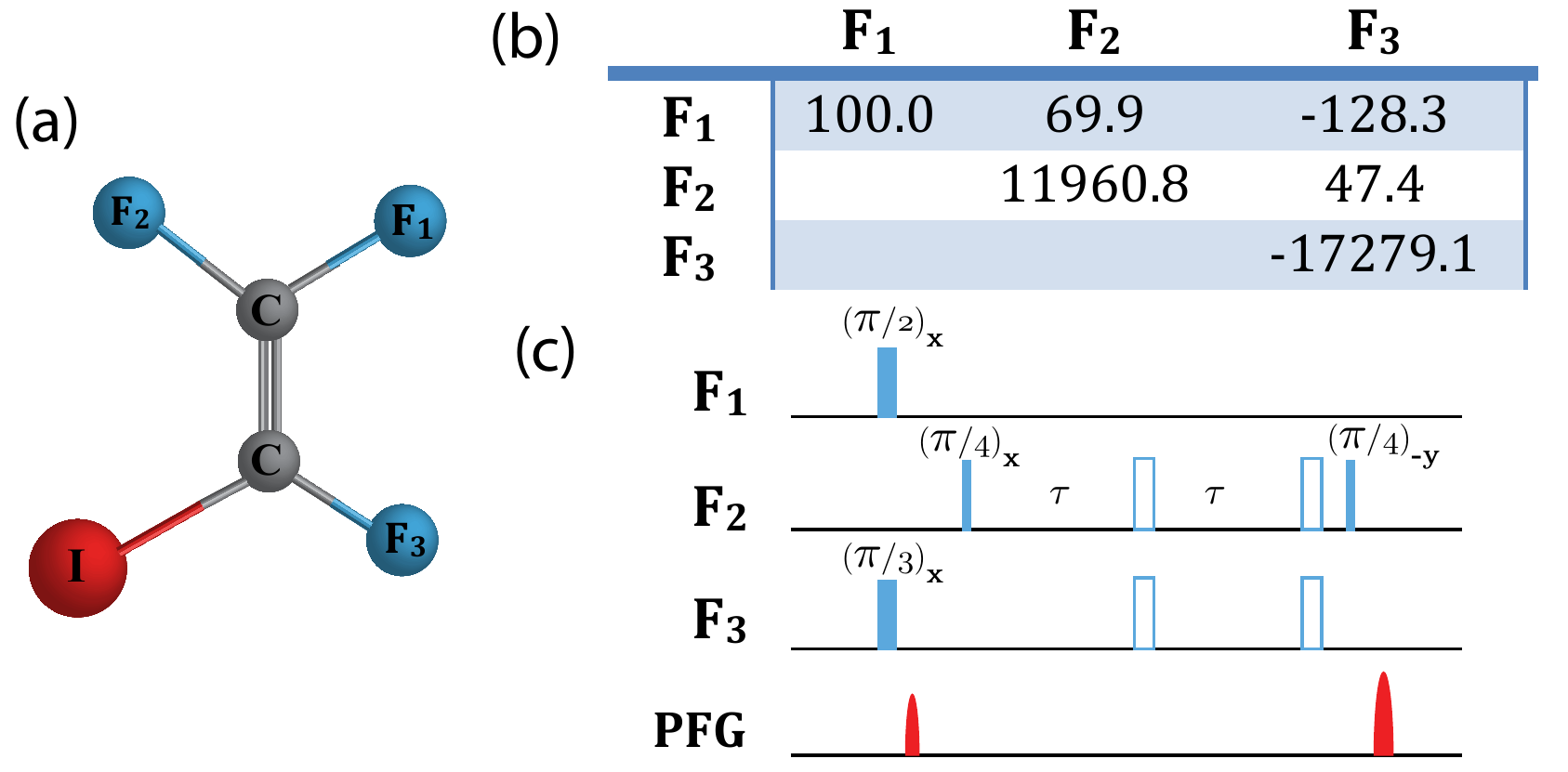}
\caption{The molecular structure of trifluoroiodoethylene (a), and corresponding chemical shifts and J-coupling values(in Hz) (b),
and the pulse sequence for the preparation of initial state (c). The open pulses are $ \pi $ pulses and 
$ \tau = 1/(4J_{23})$.}
\label{molecule}
\end{figure}

We have employed a model as shown in Fig.~\ref{ttjp} for measuring TTJP  \cite{nmr_elgi}.
The grouped gates represent the measurements in the
rotated bases.
The controlled gates shown can be either CNOT or anti-CNOT gate.
We require both CNOT and anti-CNOT gate to perform the
`ideal negative result measurement' (INRM) procedure to
measure the TTJP noninvasively, as proposed by Knee \textit{et.al.} \cite{noninvasive}. The idea behind the INRM 
procedure is as follows: consider a gate which interacts with the ancilla qubit
only when the system qubit is in state $\ket{1}$. By application of such a gate we can 
noninvasively obtain the probability of the measurement outcomes when the system was in $ \ket{0} $
state. Similarly if we have a gate, which can interact with ancilla only if the 
system qubit is in $ \ket{0} $ state then we can noninvasively obtain the probability  of the measurement
outcomes of the system state being in $ \ket{1} $. These criteria are fulfilled by the 
CNOT gate and the anti-CNOT gate respectively.

Circuit shown in Fig. \ref{ttjp} has two controlled gates for encoding the outcomes of first and second
measurements on to the first and second ancilla qubits respectively. The third measurement need not be
non-invasive since we are not concerned with the further time evolution of the system.
A set of four experiments are to be performed, with following arrangement of first 
and second controlled gates for measurement of the TTJP:
{(i)} CNOT; CNOT,
{(ii)} anti-CNOT; CNOT,
{(iii)} CNOT; anti-CNOT, and
{(iv)} anti-CNOT; anti-CNOT.

The propagators $\hat{U}_i=e^{-i \sigma_x \omega t_i/2}$ 
is realized by the cascade $\mathbb{H} \hat{U}_d \mathbb{H}$, where $\mathbb{H}$ is the Hadamard gate, and the delay propagator $\hat{U}_d = e^{-i \sigma_z \omega t_i/2}$ corresponds to the z-precession
of the system qubit at $\omega = 2 \pi100$ rad/s resonance off-set. The diagonal tomography was performed at the end to determine the probabilities \cite{nmr_elgi}. 

\begin{figure}[b]
\centering
\includegraphics[width=5cm]{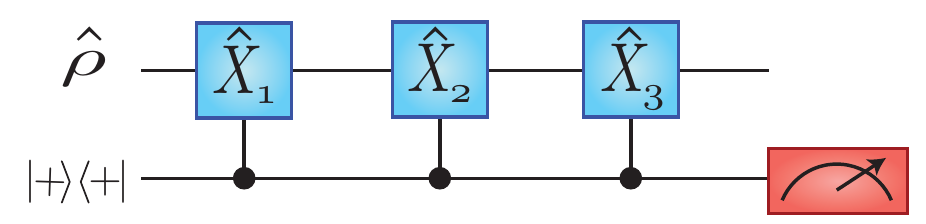}
\caption{Moussa Protocol for obtaining the $3$-time correlated moments. 
One and two time moments can be calculated using the appropriate number of controlled gates.}
\label{moments}
\end{figure}

\begin{figure}
\centering
\hspace{-.4cm}
\includegraphics[width=8.5cm]{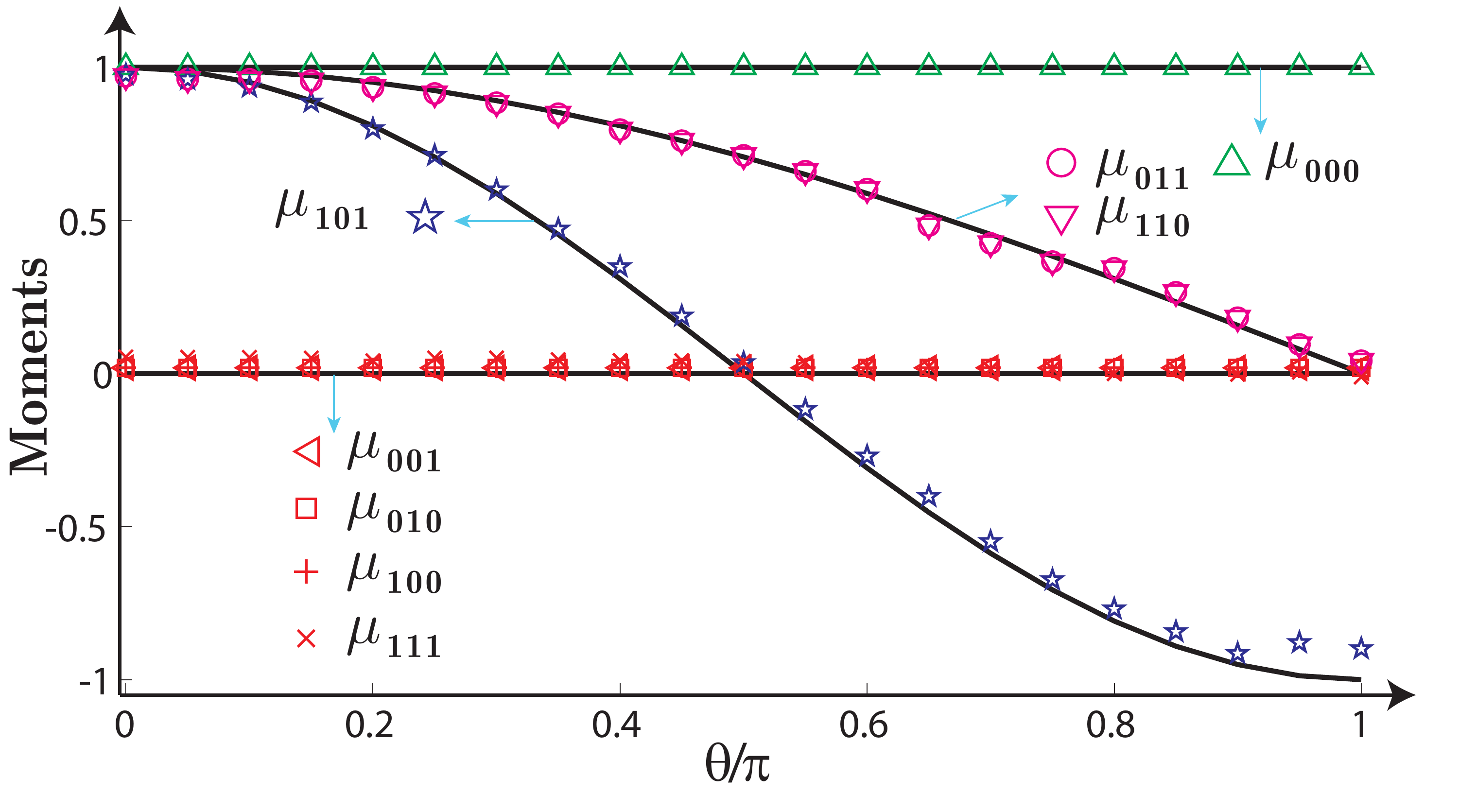}
\caption{Moments obtained experimentally from Moussa Protocol. The symbols represent experimentally obtained
values of the indicated moments with the solid lines showing the corresponding theoretical values.}
\label{moments_results}
\end{figure}

The three qubits were provided by the three $^{19}$F nuclear spins of 
trifluoroiodoethylene dissolved in acetone-D6. The structure of the molecule is shown in Fig. \ref{molecule}(a) and the chemical shifts and the scalar coupling values (in Hz) in Fig. \ref{molecule}(b).
The effective $^{19}$F spin-lattice (T$ _2^* $) and spin-spin (T$ _1 $) relaxation time
constants  were about $ 0.8 $s and $ 6.3 $ s respectively. The
experiments were carried out at an ambient temperature
of 290 K on a 500 MHz Bruker UltraShield NMR spectrometer.
The first spin (F$_1$) is used as the system qubit and, other spins (F$_2$ and F$_3$) as the ancilla qubits. 
Initialization involved preparing the state,
$\frac{1-\epsilon}{8}\mathbbm{1} + \epsilon \left\{\frac{1}{2}\mathbbm{1}_S \otimes \outpr{00}{00}_A \right\}$ where
$\epsilon \sim 10^{-5}$ is the purity factor \cite{cory}. The pulse sequence to prepare this state from the equilibrium state is shown in Fig. \ref{molecule}(c). All pulses were numerically optimized using the GRAPE 
technique \cite{Khaneja} and had fidelities better than $ 0.999 $.

\begin{figure}
\centering
\vspace{.3cm}
\includegraphics[width=8.7cm]{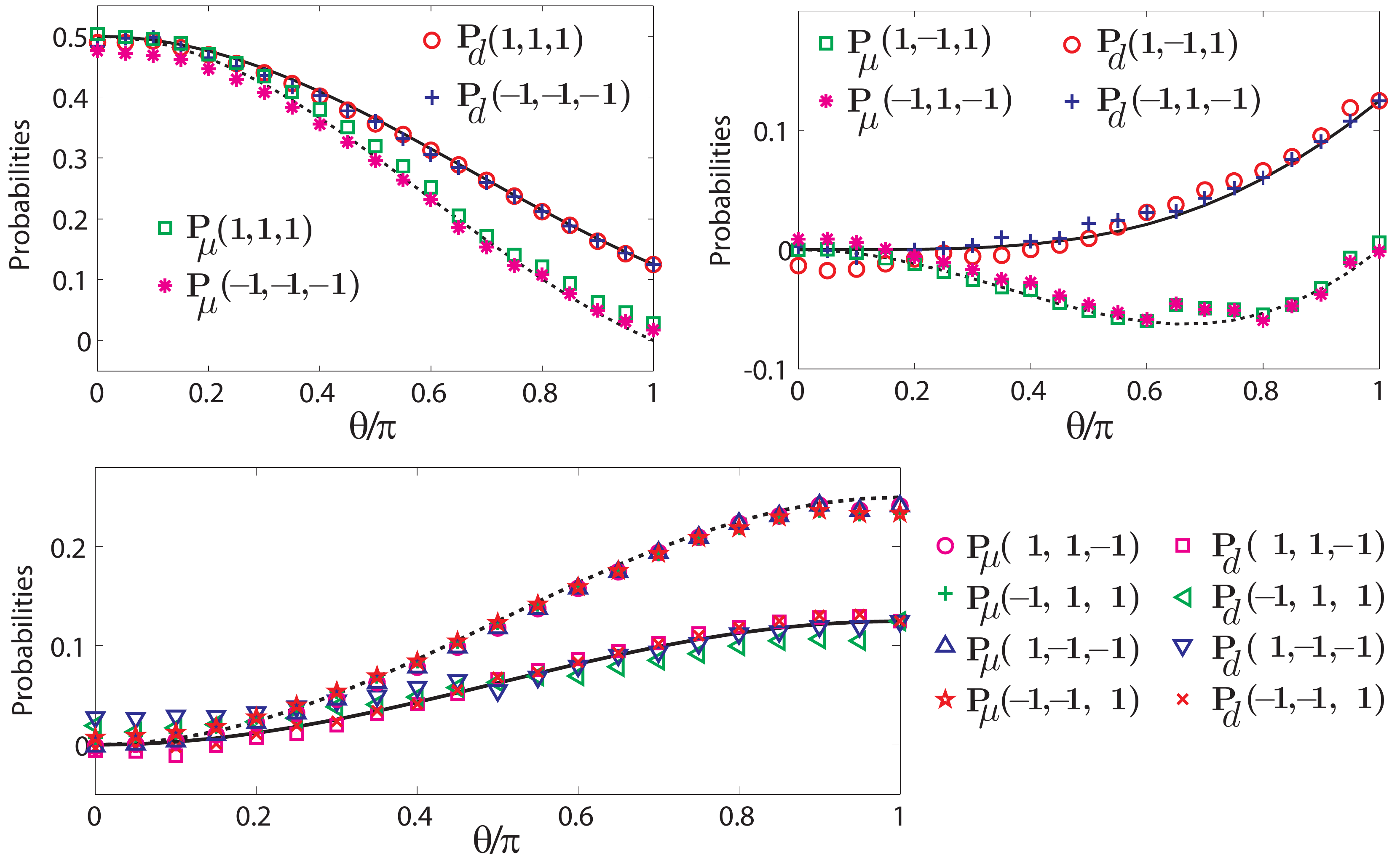}
\caption{Three-time joint probabilities (TTJP): The solid curve represents the probabilities obtained
directly and the dashed curve the probabilities obtained by inverting the moments.
The symbols represents the experimental data.}
\label{3d}
\end{figure}

With our choice of measurement model (Fig.~\ref{ttjp}) we find a striking agreement with theoretical results on TTJP (\ref{pdirect}) . One might have also run the post measured state resulting after the first dashed block in Fig.~\ref{ttjp} through an arbitrary CP map before the next step. However such post processing CP map would have affected the results. In other words, our measurement scheme provides an optimal procedure to preserve the state information, thus resulting in an excellent agreement of experimental results on TTJP with theoretical prediction (see Fig.~\ref{3d}).

For calculating the moments we utilize the Moussa protocol \cite{Moussa}, 
which requires only two spins in our case. We utilize F$_1$ as the system 
and F$_2$ as the ancilla qubit. F$_3$ was decoupled using $ \pi $ pulses 
and the initialization involved preparing the state,
$\frac{1-\epsilon}{8}\mathbbm{1} + \epsilon \left\{\frac{1}{2}\mathbbm{1}_S \otimes \outpr{+}{+}_A 
\otimes \outpr{0}{0} \right\}$, which is obtained by applying the Hadamard gate to F$_2$ 
after the pulse sequence shown in Fig. \ref{molecule}(c).
The circuit for measuring moments by Moussa protocol is shown in Fig. \ref{moments}
and it proceeds as follows, 
\begin{eqnarray}
&\hat{\rho}\otimes\ket{+}\bra{+} & \nonumber \\
&\downarrow \mathrm{c\hat{X}_1} & \nonumber \\
&\hat{\rho}\mathrm{\hat{X}_1}^\dagger\otimes\outpr{0}{1}  + \mathrm{\hat{X}_1}\hat{\rho}\otimes\outpr{1}{0} + & \nonumber \\
&\hat{\rho}\otimes\outpr{0}{0} + \mathrm{\hat{X}_1}\hat{\rho}\mathrm{\hat{X}_1}^\dagger\otimes\outpr{1}{1}& \nonumber \\
& \downarrow \mathrm{c\hat{X}_2} & \nonumber \\
&\hat{\rho}\mathrm{\hat{X}_1}^\dagger\mathrm{\hat{X}_2}^\dagger\otimes\outpr{0}{1} + \mathrm{\hat{X}_2}\mathrm{\hat{X}_1}\hat{\rho}\otimes\outpr{1}{0} + & \nonumber \\
&\hat{\rho}\otimes\outpr{0}{0} + \mathrm{\hat{X}_2}\mathrm{\hat{X}_1}\rho\mathrm{\hat{X}_1}^\dagger \mathrm{\hat{X}_2}^\dagger\otimes\outpr{1}{1}& \nonumber \\
& \downarrow \mathrm{c\hat{X}_3} & \nonumber \\
&\hat{\rho}\mathrm{\hat{X}_1}^\dagger\mathrm{\hat{X}_2}^\dagger\mathrm{\hat{X}_3}^\dagger\otimes\outpr{0}{1} + \mathrm{\hat{X}_3}\mathrm{\hat{X}_2}\mathrm{\hat{X}_1}\hat{\rho}\otimes\outpr{1}{0} + & \nonumber \\
&\hat{\rho}\otimes\outpr{0}{0} + \mathrm{\hat{X}_3}\mathrm{\hat{X}_2}\mathrm{\hat{X}_1}\hat{\rho}\mathrm{\hat{X}_1}^\dagger \mathrm{\hat{X}_2}^\dagger\mathrm{\hat{X}_3}^\dagger\otimes\outpr{1}{1},& \nonumber
\end{eqnarray}
where, $ \mathrm{c\hat{X}_i} $ represents the controlled gates and $ \hat{\rho} $ is the initial state of the system.
The state of the ancilla qubit ($ \hat{\rho}_a $) at the end of the circuit is given by,
\begin{eqnarray}
\hat{\rho}_a  &=& \outpr{0}{1}{\rm Tr}(\hat{\rho}  \hat{X}_1^\dagger \hat{X}_2^\dagger \hat{X}_3^\dagger) + \outpr{1}{0}{\rm Tr}(\hat{X}_3 \hat{X}_2 \hat{X}_1 \hat{\rho}) \nonumber \\
&&+ \outpr{0}{0}{\rm Tr}(\hat{\rho}) + \outpr{1}{1}{\rm Tr}( \hat{X}_3 \hat{X}_2 \hat{X}_1 \hat{\rho} \hat{X}_1^\dagger \hat{X}_2^\dagger \hat{X}_3^\dagger). \nonumber
\end{eqnarray}
Moussa protocol was originally proposed for commutating observables, however, it can be easily extended to non-commutating
observables. The NMR measurements correspond to the expectation values of spin angular momentum operators $ I_x $ or $ I_y $\cite{cavanagh}.
The measurement of the $ I_x $ for ancilla qubit at the end of the circuit gives: 
\begin{equation}
{\rm Tr}[\hat{\rho}_a I_x] = {\rm Tr}[\hat{X}_3 \hat{X}_2 \hat{X}_1 \hat{\rho}]/2 + {\rm Tr}[\hat{\rho} \hat{X}_1^\dagger \hat{X}_2^\dagger \hat{X}_3^\dagger]/2. 
\label{exp_Ix}
\end{equation}
If, $ \hat{X}_1, \hat{X}_2, \hat{X}_3 $
commute, then the above expression gives $ \mathrm{Tr}[\hat{\rho} \hat{X}_1 \hat{X}_2 \hat{X}_3] $.
In case of non-commuting hermitian observables, we also measure expectation value of $ I_y $, which gives :
\begin{equation}
i{\rm Tr}[\rho_a I_y] = {\rm Tr}[\hat{X}_3 \hat{X}_2 \hat{X}_1 \hat{\rho}]/2 - {\rm Tr}[\hat{\rho} \hat{X}_1^\dagger \hat{X}_2^\dagger \hat{X}_3^\dagger]/2. 
\label{exp_Iy}
\end{equation}
From (\ref{exp_Ix}) and (\ref{exp_Iy}) we can calculate  
$ \mathrm{Tr}[\hat{\rho} \hat{X}_1 \hat{X}_2 \hat{X}_3] \equiv \langle \hat{X}_1 \hat{X}_2 \hat{X}_3 \rangle$
for the $3$-measurement case. Hence, by using the different number of controlled gates in appropriate order we can calculate all the moments.
The experimentally obtained moments are shown in Fig. \ref{moments_results}.

These experimentally obtained moments are inverted according to Eq. (\ref{MI}) to calculate
the TTJP and are plotted along with the directly obtained TTJP using circuit shown in Fig. \ref{ttjp} as symbols in Fig. \ref{3d}. The theoretical values for TTJP from moments and the one directly obtained are plotted as solid and dashed lines respectively. The results agree with the predictions of Eqs.
(\ref{mip}) and (\ref{pdirect}) that the TTJP obtained directly and the one obtained from the inversion of moments do not agree.

\section{Conclusion}

In classical probability setting, statistical moments associated with dichotomic random variables determine the probabilities uniquely. When the same issue is explored in the quantum context -- with random variables replaced by Hermitian observables (which are in general non-commuting) and  the statistical outcomes of observables in sequential measurements are considered  -- it is shown that the joint probabilities do not agree with the ones inverted from the moments. This is explicitly illustrated by considering sequential measurements of a dynamical variable at three different times in the  specific example of a spin-1/2 system. An  experimental investigation based on NMR methods, where moments and the joint probabilities are extracted independently, demonstrates the moment indeterminacy of probabilities, concordant with theoretical observations.     

The failure to revert joint probability distribution from its moments points towards its inherent incompatibility with the family of all marginals. In turn, the moment indeterminacy reveals the absence of a legitimate joint probability distribution compatible with the set of all marginal distributions  -- a common underpinning of various no-go theorems in the foundational aspects of quantum theory.    

\section*{Acknowledgements}
The authors are grateful to K. R. Koteswara Rao and Sudha for discussions.
This work was partly supported by the DST project SR/S2/LOP-0017/2009.

\bibliographystyle{apsrev4-1}
\bibliography{bib_mi}

\begin{thebibliography}{10}%
\makeatletter
\providecommand \@ifxundefined [1]{%
 \ifx #1\undefined \expandafter \@firstoftwo
 \else \expandafter \@secondoftwo
\fi
}%
\providecommand \@ifnum [1]{%
 \ifnum #1\expandafter \@firstoftwo
 \else \expandafter \@secondoftwo
\fi
}%
\providecommand \enquote [1]{``#1''}%
\providecommand \bibnamefont  [1]{#1}%
\providecommand \bibfnamefont [1]{#1}%
\providecommand \citenamefont [1]{#1}%
\providecommand\href[0]{\@sanitize\@href}%
\providecommand\@href[1]{\endgroup\@@startlink{#1}\endgroup\@@href}%
\providecommand\@@href[1]{#1\@@endlink}%
\providecommand \@sanitize [0]{\begingroup\catcode`\&12\catcode`\#12\relax}%
\@ifxundefined \pdfoutput {\@firstoftwo}{%
 \@ifnum{\z@=\pdfoutput}{\@firstoftwo}{\@secondoftwo}%
}{%
 \providecommand\@@startlink[1]{\leavevmode\special{html:<a href="#1">}}%
 \providecommand\@@endlink[0]{\special{html:</a>}}%
}{%
 \providecommand\@@startlink[1]{%
  \leavevmode
  \pdfstartlink
   attr{/Border[0 0 1 ]/H/I/C[0 1 1]}%
   user{/Subtype/Link/A<</Type/Action/S/URI/URI(#1)>>}%
  \relax
 }%
 \providecommand\@@endlink[0]{\pdfendlink}%
}%
\providecommand \url  [0]{\begingroup\@sanitize \@url }%
\providecommand \@url [1]{\endgroup\@href {#1}{\urlprefix}}%
\providecommand \urlprefix [0]{URL }%
\providecommand \Eprint[0]{\href }%
\@ifxundefined \urlstyle {%
  \providecommand \doi [1]{doi:\discretionary{}{}{}#1}%
}{%
  \providecommand \doi [0]{doi:\discretionary{}{}{}\begingroup
  \urlstyle{rm}\Url }%
}%
\providecommand \doibase [0]{http://dx.doi.org/}%
\providecommand \Doi[1]{\href{\doibase#1}}%
\providecommand \bibAnnote [3]{%
  \BibitemShut{#1}%
  \begin{quotation}\noindent
    \textsc{Key:}\ #2\\\textsc{Annotation:}\ #3%
  \end{quotation}%
}%
\providecommand \bibAnnoteFile [2]{%
  \IfFileExists{#2}{\bibAnnote {#1} {#2} {\input{#2}}}{}%
}%
\providecommand \typeout [0]{\immediate \write \m@ne }%
\providecommand \selectlanguage [0]{\@gobble}%
\providecommand \bibinfo [0]{\@secondoftwo}%
\providecommand \bibfield [0]{\@secondoftwo}%
\providecommand \translation [1]{[#1]}%
\providecommand \BibitemOpen[0]{}%
\providecommand \bibitemStop [0]{}%
\providecommand \bibitemNoStop [0]{.\EOS\space}%
\providecommand \EOS [0]{\spacefactor3000\relax}%
\providecommand \BibitemShut [1]{\csname bibitem#1\endcsname}%
\bibitem{Tamarkin}%
  \BibitemOpen
  \bibfield{author}{%
  \bibinfo {author} {\bibfnamefont{J.~A.}\ \bibnamefont{Shohat}}\ and\ \bibinfo
  {author} {\bibfnamefont{J.~D.}\ \bibnamefont{Tamarkin}},\ }%
  \emph{\bibinfo {title} {The Problem of Moments}}\ (\bibinfo {publisher}
  {American Mathematical Society, New York},\ \bibinfo {year} {1943})%
  \bibAnnoteFile{NoStop}{Tamarkin}%
\bibitem{Akhiezer}%
  \BibitemOpen
  \bibfield{author}{%
  \bibinfo {author} {\bibfnamefont{N.~I.}\ \bibnamefont{Akhiezer}},\ }%
  \emph{\bibinfo {title} {The classical moment problem}}\ (\bibinfo {publisher}
  {Hafner Publishing Co., New York},\ \bibinfo {year} {1965})%
  \bibAnnoteFile{NoStop}{Akhiezer}%
\bibitem{Shi}%
  \BibitemOpen
  \bibfield{author}{%
  \bibinfo {author} {\bibfnamefont{A.~N.}\ \bibnamefont{Shiryaev}},\ }%
  \emph{\bibinfo {title} {Probability}}\ (\bibinfo {publisher}
  {Springer-Verlag, New York,},\ \bibinfo {year} {1996})%
  \bibAnnoteFile{NoStop}{Shi}%
\bibitem{note1}%
  \BibitemOpen
  \bibinfo {note} {The normalization condition $\sum_{q_1,q_2,\ldots, q_k=\pm
  1} P(q_1, q_2,\ldots q_k)=1$ is reflected in the zeroth order moment
  $\mu_{00\ldots 0}=1$}%
  \bibAnnoteFile{NoStop}{note1}%
\bibitem{nmr_elgi}%
  \BibitemOpen
  \bibfield{author}{%
  \bibinfo {author} {\bibfnamefont{H.}~\bibnamefont{Katiyar}}, \bibinfo
  {author} {\bibfnamefont{A.}~\bibnamefont{Shukla}}, \bibinfo {author}
  {\bibfnamefont{R.~K.}\ \bibnamefont{Rao}},\ and\ \bibinfo {author}
  {\bibfnamefont{T.~S.}\ \bibnamefont{Mahesh}},\ }%
  \bibfield{journal}{%
  \Doi{10.1103/PhysRevA.87.052102}{\bibinfo {journal} {Phys. Rev. A}}\ }%
  \textbf{\bibinfo {volume} {87}},\ \bibinfo {pages} {052102} (\bibinfo {year}
  {2013})%
  \bibAnnoteFile{NoStop}{nmr_elgi}%
\bibitem{Moussa}%
  \BibitemOpen
  \bibfield{author}{%
  \bibinfo {author} {\bibfnamefont{O.}~\bibnamefont{Moussa}}, \bibinfo {author}
  {\bibfnamefont{C.~A.}\ \bibnamefont{Ryan}}, \bibinfo {author}
  {\bibfnamefont{D.~G.}\ \bibnamefont{Cory}},\ and\ \bibinfo {author}
  {\bibfnamefont{R.}~\bibnamefont{Laflamme}},\ }%
  \bibfield{journal}{%
  \Doi{10.1103/PhysRevLett.104.160501}{\bibinfo {journal} {Phys. Rev. Lett.}}\
  }%
  \textbf{\bibinfo {volume} {104}},\ \bibinfo {pages} {160501} (\bibinfo
  {month} {Apr}\ \bibinfo {year} {2010}),\
  \url{http://link.aps.org/doi/10.1103/PhysRevLett.104.160501}%
  \bibAnnoteFile{NoStop}{Moussa}%
\bibitem{Fine}%
  \BibitemOpen
  \bibfield{author}{%
  \bibinfo {author} {\bibfnamefont{A.}~\bibnamefont{Fine}},\ }%
  \bibfield{journal}{%
  \Doi{10.1103/PhysRevLett.48.291}{\bibinfo {journal} {Phys. Rev. Lett.}}\ }%
  \textbf{\bibinfo {volume} {48}},\ \bibinfo {pages} {291} (\bibinfo {month}
  {Feb}\ \bibinfo {year} {1982}),\
  \url{http://link.aps.org/doi/10.1103/PhysRevLett.48.291}%
  \bibAnnoteFile{NoStop}{Fine}%
\bibitem{Bell}%
  \BibitemOpen
  \bibfield{author}{%
  \bibinfo {author} {\bibfnamefont{J.~S.}\ \bibnamefont{Bell}},\ }%
  \bibfield{journal}{%
  \Doi{10.1103/RevModPhys.38.447}{\bibinfo {journal} {Rev. Mod. Phys.}}\ }%
  \textbf{\bibinfo {volume} {38}},\ \bibinfo {pages} {447} (\bibinfo {month}
  {Jul}\ \bibinfo {year} {1966}),\
  \url{http://link.aps.org/doi/10.1103/RevModPhys.38.447}%
  \bibAnnoteFile{NoStop}{Bell}%
\bibitem{KS}%
  \BibitemOpen
  \bibfield{author}{%
  \bibinfo {author} {\bibfnamefont{S.}~\bibnamefont{Kochen}}\ and\ \bibinfo
  {author} {\bibfnamefont{E.~P.}\ \bibnamefont{Specker}},\ }%
  \bibfield{journal}{%
  \bibinfo {journal} {J. Math. Mech.}\ }%
  \textbf{\bibinfo {volume} {17}},\ \bibinfo {pages} {59} (\bibinfo {year}
  {1967})%
  \bibAnnoteFile{NoStop}{KS}%
\bibitem{Peres}%
  \BibitemOpen
  \bibfield{author}{%
  \bibinfo {author} {\bibfnamefont{A.}~\bibnamefont{Peres}},\ }%
  \bibfield{journal}{%
  \bibinfo {journal} {Journal of Physics A: Mathematical and General}\ }%
  \textbf{\bibinfo {volume} {24}},\ \bibinfo {pages} {L175} (\bibinfo {year}
  {1991}),\ \url{http://stacks.iop.org/0305-4470/24/i=4/a=003}%
  \bibAnnoteFile{NoStop}{Peres}%
\bibitem{Mermin}%
  \BibitemOpen
  \bibfield{author}{%
  \bibinfo {author} {\bibfnamefont{N.~D.}\ \bibnamefont{Mermin}},\ }%
  \bibfield{journal}{%
  \Doi{10.1103/PhysRevLett.65.3373}{\bibinfo {journal} {Phys. Rev. Lett.}}\ }%
  \textbf{\bibinfo {volume} {65}},\ \bibinfo {pages} {3373} (\bibinfo {month}
  {Dec}\ \bibinfo {year} {1990}),\
  \url{http://link.aps.org/doi/10.1103/PhysRevLett.65.3373}%
  \bibAnnoteFile{NoStop}{Mermin}%
\bibitem{LG}%
  \BibitemOpen
  \bibfield{author}{%
  \bibinfo {author} {\bibfnamefont{A.~J.}\ \bibnamefont{Leggett}}\ and\
  \bibinfo {author} {\bibfnamefont{A.}~\bibnamefont{Garg}},\ }%
  \bibfield{journal}{%
  \bibinfo {journal} {Phys. Rev. Lett.}\ }%
  \textbf{\bibinfo {volume} {54}},\ \bibinfo {pages} {857} (\bibinfo {year}
  {1985}),\ \url{http://prl.aps.org/abstract/PRL/v54/i9/p857_1}%
  \bibAnnoteFile{NoStop}{LG}%
\bibitem{UKSR}%
  \BibitemOpen
  \bibfield{author}{%
  \bibinfo {author} {\bibfnamefont{A.~R.~U.}\ \bibnamefont{Devi}}, \bibinfo
  {author} {\bibfnamefont{H.~S.}\ \bibnamefont{Karthik}}, \bibinfo {author}
  {\bibnamefont{Sudha}},\ and\ \bibinfo {author} {\bibfnamefont{A.~K.}\
  \bibnamefont{Rajagopal}},\ }%
  \bibfield{journal}{%
  \Doi{10.1103/PhysRevA.87.052103}{\bibinfo {journal} {Phys. Rev. A}}\ }%
  \textbf{\bibinfo {volume} {87}},\ \bibinfo {pages} {052103} (\bibinfo {year}
  {2013})%
  \bibAnnoteFile{NoStop}{UKSR}%
\bibitem{TP2013}%
  \BibitemOpen
  \bibfield{author}{%
  \bibinfo {author} {\bibfnamefont{M.}~\bibnamefont{Markiewicz}}, \bibinfo
  {author} {\bibfnamefont{P.}~\bibnamefont{Kurzynski}}, \bibinfo {author}
  {\bibfnamefont{J.}~\bibnamefont{Thompson}}, \bibinfo {author}
  {\bibfnamefont{S.~Y.}\ \bibnamefont{Lee}}, \bibinfo {author}
  {\bibfnamefont{A.}~\bibnamefont{Soeda}}, \bibinfo {author}
  {\bibfnamefont{T.}~\bibnamefont{Paterek}},\ and\ \bibinfo {author}
  {\bibfnamefont{D.}~\bibnamefont{Kaszlikowski}}}%
   (\bibinfo {year} {2013}),\
  \Eprint{http://arxiv.org/abs/1302.3502}{arXiv:1302.3502}%
  \bibAnnoteFile{NoStop}{TP2013}%
\bibitem{noninvasive}%
  \BibitemOpen
  \bibfield{author}{%
  \bibinfo {author} {\bibfnamefont{G.~C.}\ \bibnamefont{Knee~\textit{et.
  al.}}},\ }%
  \bibfield{journal}{%
  \Doi{10.1038/ncomms1614}{\bibinfo {journal} {Nat. Commun.}}\ }%
  \textbf{\bibinfo {volume} {3}},\ \bibinfo {pages} {606} (\bibinfo {year}
  {2012}),\ \url{http://dx.doi.org/10.1038/ncomms1614}%
  \bibAnnoteFile{NoStop}{noninvasive}%
\bibitem{cory}%
  \BibitemOpen
  \bibfield{author}{%
  \bibinfo {author} {\bibfnamefont{D.~G.}\ \bibnamefont{Cory}}, \bibinfo
  {author} {\bibfnamefont{M.~D.}\ \bibnamefont{Price}},\ and\ \bibinfo {author}
  {\bibfnamefont{T.~F.}\ \bibnamefont{Havel}},\ }%
  \bibfield{journal}{%
  \Doi{10.1016/S0167-2789(98)00046-3}{\bibinfo {journal} {Physica D}}\ }%
  \textbf{\bibinfo {volume} {120}},\ \bibinfo {pages} {82} (\bibinfo {year}
  {1998}),\ \url{http://dx.doi.org/10.1016/S0167-2789(98)00046-3}%
  \bibAnnoteFile{NoStop}{cory}%
\bibitem{Khaneja}%
  \BibitemOpen
  \bibfield{author}{%
  \bibinfo {author} {\bibfnamefont{N.}~\bibnamefont{Khaneja~\textit{et.
  al.}}},\ }%
  \bibfield{journal}{%
  \Doi{10.1016/j.jmr.2004.11.004}{\bibinfo {journal} {Journal of Magnetic
  Resonance}}\ }%
  \textbf{\bibinfo {volume} {172}},\ \bibinfo {pages} {296} (\bibinfo {year}
  {2005}),\ ISSN \bibinfo {issn} {1090-7807},\
  \url{http://www.sciencedirect.com/science/article/pii/S1090780704003696}%
  \bibAnnoteFile{NoStop}{Khaneja}%
\bibitem{cavanagh}%
  \BibitemOpen
  \bibfield{author}{%
  \bibinfo {author} {\bibfnamefont{J.}~\bibnamefont{Cavanagh}},\ }%
  \emph{\bibinfo {title} {Protein NMR spectroscopy: principles and practice}}\
  (\bibinfo {publisher} {Academic Pr},\ \bibinfo {year} {1996})%
  \bibAnnoteFile{NoStop}{cavanagh}%
\end{thebibliography}%


\end{document}